\documentstyle[12pt]{article}

\begin{document}

\noindent {\large {\bf A Critique of the Gauge Technique}}

\vspace{1in}

\noindent {\em R. Delbourgo}

\medskip

{\small \noindent School of Mathematics and Physics,
        University of Tasmania\\
        GPO Box 252-21, Hobart, Australia 7001 \vspace{1in} }

{\small \noindent {\em Abstract} }

{\small \noindent A summary of the successes of and obstacles to the gauge
technique (a non-perturbative method of solving Dyson-Schwinger equations
in gauge theories) is given, as well as an outline of how progress may be
achieved in this field.}

\medskip

\section{The gauge technique}

Nature seems inordinately fond of gauge theories. Chromodynamics,
electroweak
theory and gravitation are all based on the gauge principle, featuring the
groups SU(3), SU(2)$\times$U(1) and SL(2,c), respectively. It is quite
likely that any unified gauge model of the fundamental forces will also be
a gauge field theory and it is conceivable that a supersymmetric version,
in some higher dimension, will be founded on a local supergauge principle,
although nature seems reticent to display her supersymmetric hand in the
low energy spectrum of states.

Whenever a field theory is invariant under local group transformations,
the resulting Green functions obey a series of `gauge covariance' relations
which specify how the functions change under a variation of gauge\footnote{
Purely photonic processes and amplitudes involving on-shell fermions remain
{\em gauge-invariant} of course.}. In QED these relations were
originally found by Landau and Khalatnikov (1956, LK for short); later on,
Johnson and Zumino (1959) and Zumino (1960) rederived them using functional
methods. An immediate consequence of these ideas is the famous divergence
properties of Green functions at each vector leg, named Ward-Green-Takahashi
(WGT) identities after their early discoverers (Ward 1950; Green 1953;
Takahashi, 1957).

Heroic efforts have been expended in determining the behaviour of off-shell
Green functions. This program has be carried out to high accuracy in
perturbation theory for the standard model of particle physics and the
results (at least for asymptotic momenta in QCD) are quite reliable because
the perturbation series `converge' to the order in which they have been
calculated so far and also because the model is
renormalizable\footnote{They will eventually go wrong because the series in
$e^2$ is believed to be only an asymptotic one.}.
An alternative approach is to solve the full set of Dyson-Schwinger (DS)
equations connecting the various Green functions by some suitable
truncation procedure. For instance, one simple way is to sum the `rainbow
diagrams' of the self-energy or the `ladder diagrams' of the vertex, which
amounts to simplifying the DS equations for the two- and three-point
functions in a particular way. By this means one can already discern the
phase structure of a model as the coupling runs through certain critical
values.

In gauge theories, thanks to the existence of WGT identities between
successive $n$-point Green functions, a more sophisticated approach
suggests itself. The identities allow one to determine (longitudinal)
pieces of the Green function in terms of the function with one less vector
particle\footnote{There is even a `curly' version of such identities,
obtained by Takahashi; they are rather more complicated than the ordinary
divergence and we will have little to say about them in this paper. However
in 2-dimensions, it should be noted that they are equivalent to the axial
gauge identities (Delbourgo and Thompson 1982; Thompson 1983; Kondo 1997).}.
Hence if one discards transverse parts of the
Green function (orthogonal to the photon momentum) or at least ties them in
a particular fashion to the longitudinal parts, then this amounts to a
truncation which produces a closed set of equations among the
functions---to some level of approximation depending on $n$.
For want of better terminology, we will coin this procedure the `gauge
technique' after Salam (Salam 1963; Salam and Delbourgo 1964;
Strathdee 1964)---even though it was originally tied to a
specific way of `solving' the gauge identities and began at level $n=2$.
Anyway, armed with solutions of the DS equations, one may go on to
calculate quantities of physical interest, such as decay constants of
hadrons (Maris and Roberts 1997; Ivanov {\em et al} 1998).
This approach (Thompson and Zhang 1987; Roberts and Williams 1994)
can be carried out in any number of space-time dimensions $D$ and for any
physical fermion mass $m$ and it is quite interesting to see how the
results depend on $D$ and $m$ separately, in various background (Cornwall
1986; Maris and Roberts 1998) configurations (temperature, field, etc.).

In this paper we shall focus primarily on the archetype gauge theory, QED.
This has the advantage of yielding simple abelian identities and of
bringing into relief the principal obstacles which the gauge technique
must circumvent before it can be considered a fully-fledged method, with
results that are above reproach. The purpose of this paper is to highlight
some of these problems and to suggest possible ways in which some progress
may be achieved.

Since we are dealing with gauge theories, the first issue we face
is the gauge covariance of the technique, i.e. how well do the
Green functions solutions obey the LK relations. The second question
concerns the comparison with perturbation theory (in the fine structure
constant $\alpha$), i.e. how closely do the non-perturbative answers
coincide with the perturbation series if we try to expand them in $\alpha$.
The third matter refers to the level of approximation, i.e. to what
extent do the results change with the level of truncation $n$. Since the
technique concentrates primarily on the charged particle functions, the
fourth issue is what happens to vector propagation, i.e. how will vacuum
polarization pan out in such schemes: this is an serious matter
because photon renormalization, not electron renormalization, governs the
momentum dependence of the running charge. Indeed vertex and charged
particle renormalizations are largely irrelevant: they are gauge-dependent
and cancel out, leaving little physical trace. It means that going from
quenched to unquenched solutions represents a major step physically.

If these issues can be clarified in QED, there are definite lessons
(Papavassiliou and Cornwall 1986; Alekseev 1998; Cahill and Gunner 1998)
for QCD and gravity.

\section{Gauge identities and SD equations}
First we set up the framework for the ensuing discussion by summarising
quickly the identities and the coupled equations for Green functions in
QED. If nothing else this will fix our notation.
Let $G_{\lambda_1,\lambda_2,.}(p_1,p_2,.;k_1,k_2,.)$ stand for a
connected time-ordered Green function, with electron momenta $p$
and photon momenta $k$, as labelled. It is usual to write $G(p)=S(p)$ and
$G_{\mu\nu}(k)=D_{\mu\nu}(k)$ for the two-point electron and photon
propagators respectively. It is convenient often to define the one-particle
irreducible Green functions $\Gamma(p;k)$ by dropping all the pole parts of
$G(p;k)$ and by multiplying out by the inverse propagators of the external
lines; for instance, the `3-vertex function' $\Gamma(p'\!,p;k)$ with
$k=p'\!-\!p$ arises via $G_\lambda(p'\!,p;k)\!\equiv\!S(p'\!)\Gamma^\mu
 (p'\!,p;k)S(p)D_{\mu\lambda}(k)$, while the `Compton part'
$\Gamma(p',p;k',k)$ of electron photon scattering occurs in
\begin{eqnarray}
G_{\lambda\kappa}(p',p;k',k) &\equiv & D_{\lambda\nu}(k')S(p')
[\Gamma^{\nu\mu}(p',p;k',k)- \nonumber\\
 & & \Gamma^\nu(p',p'+k';k')S(p+k)\Gamma^\mu(p+k,p;k)- \nonumber \\
 & &\Gamma^\nu(p\!-\!k',p;k')S(p'\!-\!k)\Gamma^\mu(p',p'\!-\!k;k)]S(p)
 D_{\mu\kappa}(k)
 \end{eqnarray}
etc. The $x$-space Fourier transforms are obtained as convolutions; thus
$$G_\lambda(x,y;z)=\int d^4\!x'd^4\!y'd^4\!z'D_{\lambda\mu}(z-z')S(x-x')
  \Gamma^\mu(x',y';z')S(y'-y),$$
and so on.  The reason why we have mentioned the coordinate space
version is because the gauge covariance identities, to which we shall
presently turn, are best written in $x$-space.

The Dyson-Schwinger equations connect successive (renormalized) charged
Green functions through the series\footnote{Below and later on, we adopt
the notation, $\bar{d}^4k\equiv d^4k/(2\pi)^4,\bar{\delta}^4(k)=
(2\pi)^4\delta^4(k)$, etc.},
\begin{equation}
S(p)(\!\not\!p-m_0) = Z^{-1}+ie^2\int\bar{d}^4\!k
  G_\lambda(p,p-k;k)\gamma^\lambda
\end{equation}
\begin{equation}
G_\mu(p',p;p'-p)(\!\not\!p-m_0)=S(p')\gamma_\mu - ie^2\int\bar{d}^4\!k
           G_{\mu\lambda}(p',p-k;p'-p,k)\gamma^\lambda
\end{equation}
etc. plus the uncharged cases starting with,
\begin{equation}
\eta_{\mu\nu} = Z_A\,{D^{-1}(k)_{0\mu}}^\lambda D_{\lambda\nu}(k)
            -ie^2Z\,\,{\rm Tr}\int\bar{d}^4\!p G_\mu(p+k,p)\gamma_\nu.
\end{equation}
The nonlinearity of electrodynamics becomes evident via the infinite
skeleton expansion of the higher-point Green functions, when one
expresses the $G$ in terms of their 1PI counterpart functions $\Gamma$.
That is why one is obliged to invoke some kind of truncation for solving
the equations nonperturbatively, instead of resorting to $e^2$
expansion---the idea being that a nonperturbative solution may reveal
some dependence in $1/e^2$ (usually in an exponent) which is not
immediately apparent from the asymptotic $e^2$ series.

All of the above applies, however one fixes the photon gauge. Now
under a change of photon gauge, $D_{\mu\nu}(x)\rightarrow D^M_{\mu\nu}(x)
= D_{\mu\nu}(x)\!-\! \partial_\mu\partial_\nu M(x)$, the renormalized Green
functions are modified in a well-defined manner from $G$ to $G^M$. Thus
$$S^M(x) = \exp[ie^2M(x)]S(x),$$
$$G^M_\mu(x,y;z)\!=\! \exp[ie^2M(x-y)][G_\mu(x,y;z)\!+\!
             iS(x-y)\partial^z_\mu\!\{M(x-z)-M(y-z)\}], $$
and so on. Let us call these the `gauge covariance' or LK relations. It is
simple to check that they are consistent with the SD equations in any gauge.
A secondary consequence (though historically a primary one) is that the
Green functions will satisfy the WGT identities,
$$(p'\!-\!p)^\lambda G_\lambda(p',p;p'-p) = S(p) - S(p'),$$
$$k^\mu G_{\mu\nu}(p',p';k',k)=G_\nu(p',p+k;k')-G_\nu(p'-k,p;k'),$$
whose soft $k$-limit produces the Ward versions,
$G_\mu(p,p;0)=-\partial S(p)/\partial p^\mu$, etc.  The WGT identities
also appear straightforward for the 1PI functions $\Gamma$, e.g.
$$(p'-p)^\lambda \Gamma_\lambda(p',p;p'-p) = S^{-1}(p') - S^{-1}(p).$$
However the WGT identities are weaker than the LK relations, which can
themselves become quite complicated for the amputated $\Gamma$, in
contrast to those for the full Green functions $G$, especially when
written in momentum space.
It is therefore a vexing business to verify that any $\Gamma$, derived
somehow in some $M$-gauge, properly obeys the required covariance
identity; by comparison it is easier to investigate the matter for the
non-amputated $G$. In fact if one only amputates the photon legs, the
gauge covariance relations simplify a little further and make our task
easier. For illustration, take the vertex function $S\Gamma S$;
one finds, like $S^M(x-y)=\exp[ie^2M(x-y)]S(x-y)$, that
\begin{equation}
(S\Gamma S)^M(x,y,;z) = \exp[ie^2M(x-y)] (S\Gamma S)(x,y;z),
\end{equation}
\begin{equation}
{\rm and}\quad \partial_z\cdot(S\Gamma S)^M(x,y;z) =
                 iS^M(x-y)[\delta^4(x-z)-\delta^4(y-z)].
\end{equation}

That completes the quick tour of the basis for the gauge technique.

\section{What gauge covariance implies}
Before tackling the spinor case, let us start with scalar electrodynamics,
where the algebra and arguments are simpler. If $\Delta$ connotes the
charged scalar propagator and $\Gamma$ the fully amputated 3-point
vertex in some gauge (specified by $M=0$ say), focus on the three relations:
\begin{eqnarray}
\Delta^M(x,y) &=& \exp[ie^2M(x-y)]\Delta(x,y),\nonumber \\
(\Delta\Gamma\Delta)^M(x,y;z) &=& \exp[ie^2M(x-y)]
        (\Delta\Gamma\Delta)(x,y;z), \\
\partial_z^\mu(\Delta\Gamma_\mu\Delta)^M(x,y;z) &=&
    i\Delta^M(x-y)[\delta^4(x-z)-\delta^4(y-z)],\nonumber
\end{eqnarray}
associated with the lowest functions in a different gauge $M$. Now in
general, the off-shell 3-point function can be expressed in terms
of two invariants, one associated with the longitudinal vertex and the
other with a purely transverse vertex ($k=p'-p$):
\begin{equation}
\Gamma_\mu(p',p;k)=(p'+p)_\mu L(p'^2,p^2,k^2) +
   [k_\mu(p'^2\!-\!p^2)-(p'+p)_\mu k^2] T(p'^2,p^2,k^2),
\end{equation}
where $L$ and $T$ are symmetric scalar function under $p^2\leftrightarrow
p'^2$. It follows that, when $p^2=p'^2$, the transverse part can be
effectively combined with the longitudinal piece; this case applies in
particular when one goes on the meson mass shell.

It is the $T$ part
which largely governs\footnote{It is worth noting that $T$ has no
singularities in the triangular variable $k^4+p^4+p'^4-2p^2p'^2-2p^2k^2-
2p'^2k^2$, but that if we set $p^2\neq p'^2$ there do arise logarithmic
divergences (Ball and Chiu 1980) in perturbation theory as $k^2\rightarrow
0$.}
the meson `form factor' because the scalar WGT identity tells us
unambiguously that the longitudinal part $L$ cannot depend on the
invariant $(p'-p)^2$ in {\em any\/} gauge; for it is {\em always\/} true
that
\begin{equation}
L^M(p'^2,p^2) = (\Delta^{-1\,M}(p')-\Delta^{-1\,M}(p))/(p'^2-p^2).
\end{equation}
Like $L$, the transverse contribution $T$ must also change with gauge
function $M$, but in a subtler way than $L$ and one where the $(p-p')^2$
dependence cannot be so easily forgotten. For suppose
that in some gauge, we were to ignore $T$ and wrote at the very least:
$$(\Delta\Gamma_\mu\Delta)(p',p;k)=\Delta(p')(p+p')_\mu\Delta(p)L. $$
Then in another gauge, according to (7), the result would get modified to
$$(\Delta\Gamma_\mu\Delta)(p',p;k)=\int \bar{d}^4\!k'
  \Delta(p'-k')(p+p'-2k')_\mu\Delta(p-k')L E^M(k'),$$
where
\begin{equation}
 E^M(k) \equiv \int d^4\!x\,\exp[ie^2M(x)+ik\cdot x].
\end{equation}
A transverse amplitude $T$ in the off-shell vertex would be ineluctably
entrained via the momentum numerator of the integrand. For instance,
starting with first order perturbation theory, a transverse
Lorentz-covariant lurks within the expression
\begin{equation}
(\Delta\Gamma_\mu\Delta)^M(p',p;k)=\int \bar{d}^4\!k'
         \frac{(p+p'-2k')_\mu E^M(k')}{[(p'-k')-m^2][(p-k')^2-m^2]}.
\end{equation}
Only in the limit $e^2M=0$, when $E^M(k')=\bar{\delta}^4\!(k')$, will
such a transverse term disappear. This argument teaches us two things
about ensuring gauge covariance for general $M$:
(i) that it is perilous to neglect the transverse parts of Green functions
in non-perturbative treatments, and
(ii) that one cannot always disregard the dependence of the amplitude
on the momentum of the {\em vector} leg, i.e. one cannot
purely\footnote{This applies with force to various `improvements' or
corrections to the longitudinal vertex, consistent with multiplicative
renormalizability, that have been suggested (Curtis and Pennington, 1993;
Haeri 1993). In this connection it is worth remembering that the covariant
gauge $a=3$ produces a zero first order in $\alpha$ correction to the scalar
propagator or to all orders, in the infrared limit.} use
functions of $p^2, p'^2$.

The discussion increases in substance, richness and delicacy for
fermions. Instead of one transverse and one longitudinal part, the
vertex contains four independent longitudinal pieces and eight
transverse pieces off-shell. In a real tour de force, these pieces
have been computed by Ball and Chiu (1980) to first order
perturbation theory in the Fermi-Feynman gauge, and by Kizilersu,
Pennington and Reenders (1995) in any gauge. Being off-shell, the
answers are extremely involved and we shall content ourselves with
making three remarks: (i) there exist transverse covariants now which
survive the on-shell spinor limit, such as $i\sigma_{\mu\nu}k^\nu$,
that have important physical consequences; (ii) no longer can one
combine transverse vertices with longitudinal ones for physical
fermions\footnote{In fact because physical fermions satisfy
free equations of motion, the transverse covariants are no longer
independent but can be transmogrified into one another.};
(iii) there is a lot more freedom in `solving' the WGT identity
for the vertex, with distinct methods, all deemed free of unphysical
singularities and all independent of the square of the photon
momentum, yielding vertices differing by specific transverse
terms $T$. (See the Appendix for details.) The various $T$ arise
automatically in the spinor version of (8) for arbitrary $M$, so it is
futile to discard them in a general gauge\footnote{Nevertheless one
should recognize the privileged position of the covariant (Landau)
gauge $a=0$, because the first order in $\alpha$ correction to the
wave function renormalization vanishes identically, as do all
rainbow modifications of the spinor propagator. See Section 6.}
unless one gives up on the idea of satisfying the LK relations---a
big disappointment for a gauge theory.

\section{Comparison with perturbation theory}
If one does succeed in obtaining `acceptable' solutions (presumably
with an implicit dependence on $e^2$) of the DS equations by the gauge
technique, then there is the obvious question of how well the answers
stack up against perturbation theory, when expanded in powers of $e^2$.
At the very least one would hope that they would correct up to first order
in $\alpha$, as they are indubitably correct to zeroth order---but that
hardly constitutes progress! It is not much good having them agree with
perturbation theory in one gauge (for some $L$ and $T$) but being wrong
in another gauge. However that is what will likely happen from the
transformation property of the $G$ {\em unless} the various parts carry
precisely the correct dependence on $M$ and, as stressed previously, one
countenances some $k^2$ dependence in $S\Gamma S$. And it is no good
avouching that the off-shell dependence of the propagator on the momentum
looks `reasonable', with `suitable behaviour' in the infrared or
ultraviolet limits, since one can change the behaviour at will, just
by choosing the gauge function $M$ however one likes.

To clarify these points, let us consider covariant gauges,
parametrized by a real number $a$:
$$D_{\mu\nu}^a(k) = -\eta_{\mu\nu}/k^2 + (1-a)k_\mu k_\nu/k^4.$$
(The values $a=0,1,3$ define the Landau, Fermi-Feynman and Fried-Yennie
gauges respectively.) Upon introducing a regularization scale, like the
electron mass $m$, the Fourier transform of $M(k) \equiv -a/k^4$ is
obtained as $iM(x)=-a\ln(-m^2x^2)/16\pi^2$. Therefore the coordinate
space Green functions $G$ of two charged fields separated by $x$ will be
multiplied by the gauge factor
$$\exp[ie^2M(x)] = (-m^2x^2)^{-a\epsilon};\qquad \epsilon \equiv
e^2/16\pi^2 = \alpha/4\pi,$$
with transform,
$$E^M(k) = \int d^4x\,\exp(ip\cdot x)(-m^2x^2)^{-a\epsilon}
         = -\frac{16\pi^2i\Gamma(2-a\epsilon)}{k^4\Gamma(a\epsilon)}
            (-k^2m^2/4)^{a\epsilon}. $$
Expanding in powers of $e^2$, one can check that
$$E^M(k) = \bar{\delta}^4(k) - ie^2a/k^4 + O(e^4),$$
with the second term on the right corresponding to the change of
covariant gauge parameter $a$ to first order perturbation theory.

The transverse term has a complicated analytic form which only simplifies
in various asymptotic limits; the behaviour in the ultraviolet regime is
one such limit and King (1983) and Haeri (1988) put it to good use in
correcting multiplicative renormalizability that is jeopardised in simple
gauge technique ansatze; however, their procedure fails in the soft photon
limit and does not do justice to the analytic behaviour in momentum
transfer.
To do better, it is useful to look at the form of $T$ in scalar
electrodynamics say, to first order in $\alpha$ or $M$ as above.
An examination of the Feynman graph integral, shows that it can be
expressed in the parametric form
\begin{eqnarray}
 (p'^2-p^2)T & = & \frac{e^2}{16\pi^2}\int d\alpha d\beta d\gamma
   \,\,\delta(1-\alpha-\beta-\gamma)\cdot \nonumber \\
 & & \frac{N(\alpha,\beta,\gamma)}
     {m^2(1-\alpha)-(p'^2\beta+p^2\gamma)\alpha-(p-p')^2\beta\gamma}
\end{eqnarray}
where the parametric numerator (for $a=1$ say) is given by
$$N = 2(p'^2-p^2)\alpha + (\beta-\gamma)[p^2(1-\beta)+p'^2(1-\gamma)
                                         +m^2(3-\alpha)] $$
$$-3(\beta-\gamma)\beta\gamma[(p'^2-p^2)\alpha+(p-p')^2(\beta-\gamma)].$$
The important thing is that the integral produces a result which vanishes
for $p^2=p'^2$.

It is interesting to see whether the answer can be rewritten as some
kind of dispersion relation. To that end, consider the self-energy first.
The standard Feynman parametric form leads to the scalar integral
$$I = \int_0^1 d\alpha \,\,\chi(p,\alpha)/(p^2\alpha-m^2).$$
By changing variable to $W^2 = m^2/\alpha$, the expression above can be
easily converted into the more familiar form
$$I =\int_{m^2}^\infty \rho(W^2)/(p^2-W^2).$$
Turning to the proper vertex part, if we make the change of variables in
the Feynman parametric integral, $\beta+\gamma=\sigma, \beta-\gamma=\sigma
u$ and $\sigma=m^2/W^2$, we obtain an expression like
$$\int_{m^2}^\infty dW^2 \int_{-1}^1 du\,\, N(p',p,u,W)/D(p',p,u,W),$$
for the transverse part. The denominator $D$ takes the form
$$p'^2(1+u)/2+p^2(1-u)/2+(p'-p)^2(1-u^2)(1-m^2/W^2)(1-u^2)/4-W^2,$$
and we see that only near the fermion mass shell ($W^2=m^2$) and
in the soft photon limit can one disregard the dependence on the momentum
transfer. (In the ultraviolet limit where the spinor momentum is a long
way from the $p^2=m^2$ one can make other sorts of approximations.)

What this exercise demonstrates is that under no stretch of the imagination
can one invoke a transverse vertex which is purely a function of $p^2$ and
$p'^2$. Even an invokation like
$$(S\Gamma^T_\mu S)(p',p)=\int dW
        \frac{1}{\not{p}'-W}i\sigma_{\mu\nu}(p'-p)^\nu\frac{1}{\not{p}-W},$$
misses the point altogether, since it fails to include the correct analytic
structure for a magnetic {\em form factor}. Remember also that the
covariance LK relation under change of gauge will inevitably create such
structure, if none were initially present. Therefore we insist that any
substantive improvements to the gauge technique ought to include these kinds
of effects and agree with first order perturbation theory at the very least.
Ansatze of the type,
$$\Gamma^T_\mu(p',p)=\int_m^\infty dW\int_{-1}^1 du\,\,
         \rho^T(W)K^T_\mu/D(p',p,u,W)$$
stand some chance of capturing the main feature of transverse corrections.

\section{Level of truncation}
In an effort to improve upon the three-point Green function it is good to
proceed to the next level of the DS equations. This will relate the three
and four-point Green functions, via a Bethe-Salpeter like equation---see
Parker (1984), Delbourgo and Zhang (1984).
Note that in order to make use of the higher-point WGT
identities, it is sensible to write this equation with one of the spinor
legs preferred rather than the photon leg as is normally done; for with that
choice one can use the four-point WGT identity to relate to relate that
amplitude to the three-point Green function and thereby arrive at a
self-consistent equation for the three-point function. (This is in complete
analogy to the favourite way of handling the propagator and vertex function
together and we can think of it as the $n=3$ level improvement of the gauge
technique, in contrast to the conventional $n=2$ level).

What is more to the point is that, at this level, the longitudinal and
transverse vertices are treated {\em on an equal footing} while the
two-point function (propagator) is obtained secondarily through the
divergence
of the three-point function.  In practice, though, the equation for the
vertex is {\em very} difficult to solve, for one is not entirely sure of its
{\em full} analytic representation, except through the perturbation
expansion.
The best one can do in such circumstances is to use a double dispersion
relation (in a Feynman parametric form say) and try to determine the
spectral
function self-consistently, but even that is not straightforward. The only
progress to date has been the determination (Delbourgo and Zhang 1984) of
the
propagator spectral function in a manner which coincides with perturbation
theory to order $\alpha^2$. Hopefully this problem will receive due
attention
in future.

\section{Unquenching -- vacuum polarization}
In most of the self-consistent calculations of the charged field propagator,
the photon is taken as bare: this represents the so-called ``quenched
approximation''. Including the effect of vacuum polarization leads to a
highly nonlinear equation for the propagator even in the best method of
solving for the propagator, since vacuum polarization is determined by
charged loop effects; that is one of the main reasons why the problem
is normally avoided.  The best that one can do in that situation is to make
an inspired guess at the behaviour of the dressed vector propagator $D(k)$
and (i) either include this in the computation of $S(p)$, or (ii) introduce
a running coupling in the interaction between the vector and the charged
field.  Now in QED the photon receives {\em small} logarithmic corrections
in the ultraviolet regime, but its low energy properties are largely
unaffected; for that reason the quenched approximation is not too drastic
a procedure for electrodynamics, at least in four dimensions\footnote{
In three dimensions charged loops alter substantially the low energy
behaviour
from $D(k) = 1/k^2$ to $D(k) \propto 1\sqrt{-k^2}$, while in two dimensions
the vector becomes massive through the Schwinger mechanism.}.
But in QCD, the gluon exhibits asymptotic freedom in the ultraviolet---again
a logarithmic correction---which hints at infrared slavery (confinement
of colour?) at low energies and a propagator which is possibly more
singular that the undressed form (Alekseev 1998; Cahill and Gunner 1998).
Thus a variety of models have been proposed and the corresponding quark
$S(p)$
found; the results depend critically on the assumed form of $D(k)$ which is
itself influenced by the gluon self-interactions and the ghost field
effects.
There is still some dispute about what is the correct form of $D(k)$ and
whether the quark propagator is an entire function of the momentum. We do
not
wish to get involved in these arguments; suffice it to say that the method
with
the best chance of being correct is the one that handles the gauge
covariance
properties correctly, in tandem with the gauge-invariant effective coupling
$g = g_0Z_1^{-1}Z_2Z_3^{1/2}$.

In this connection it is worth recalling (Larin and Vermaseren 1993)
the perturbative expansions of the beta functions for QED and QCD,
which are of course gauge independent.
Let $\epsilon \equiv g^2/16\pi^2 \rightarrow \alpha/4\pi$ for QED. Then
$$\beta(\epsilon)=\epsilon[\gamma_3(\epsilon)+2\gamma_2(\epsilon)-2\gamma_1
(\epsilon)] \equiv \sum_{n=1}^\infty \beta_n \epsilon^{n+1}$$
(In QED the anomalous dimensions are equal, $\gamma_2=\gamma_1$, so $\beta
=\epsilon\gamma_3$ is determined purely by the anomalous dimension of the
photon field.) Up to three loops,
\begin{eqnarray}
\beta_{{\rm QED}}\!\!\!&=\!\!\!&\frac{4}{3}\epsilon^2\left[ 1 + 3\epsilon-
        (\frac{3}{2}+\frac{11N}{3})\epsilon^2 + \ldots\right]\\
\beta_{{\rm QCD}}\!\!\!&=&\!\!\!\epsilon^2\left[(\frac{2N}{3}-11)\!+\!
       (\frac{38N}{3}-102)\epsilon\!-\!(\frac{325N^2}{54}-\frac{5033N}{18}
       +\frac{2857}{2})\epsilon^2\!+\!\ldots\right]\nonumber,
\end{eqnarray}
where $N$ stands for the number of charged fermions or flavours. The
dependence
on the gauge parameter $a$ in QED arises through the anomalous scaling
function
for the spinor\footnote{See Larin and Vermaseren (1993) for a complete
determination of the anomalous scaling functions in QCD, which are too long
to
reproduce for the purposes of this paper.},
$$\gamma_2(\epsilon)= \epsilon\left[ a - \frac{3}{2}\epsilon + \frac{3}{2}
                      \epsilon^2 + \ldots \right].$$
Notice that the gauge dependence arises only at one-loop level and that it
vanishes in the Landau gauge $a=0$. More significantly there are higher
order
in $\alpha$ contributions which cannot be ignored; it is therefore fatuous
to
suppose that one can simply set the coefficient of $\not{p}$ in the spinor
propagator equal to one in the Landau gauge---this is simply incorrect in
higher
orders.

Anyhow it is reasonable to enquire what repercussions, if any, does the
gauge
technique have on vacuum polarization. There are two important points to
check
in this connection: whether the polarization tensor $\Pi_{\mu\nu}(k)$ is
transverse and whether it is gauge-independent. Since
$$\Pi_{\mu\nu}(k)=ie^2\,{\rm Tr}\int\bar{d}^4p(S\Gamma_\mu S)(p+k,p)
        \gamma_\nu,$$
it is fairly clear that a non-perturbative calculation of $\Pi$ which
dresses
the fermions ($S$) but leaves the full vertex as bare
($\Gamma\rightarrow\gamma$),
will not produce a transverse tensor; therefore this strategy is
unacceptable.
It is also easy
to concoct a longitudinal approximation to the three-point Green function
that
{\em does} lead to transverse polarization, e.g. a mass-weighted spectral
representation, as described in the appendix. Carrying out the computations
in the Landau gauge say, where
$$ S(x-y) = \int dW\,\rho_0(W)S_F(x-y|W),$$
\begin{equation}
 (S\Gamma^L S)(x,y;z)=\int dW\,\rho_0(W)S_F(x-z|W)\gamma S_F(z-y|W),
\end{equation}
with $S_F(x|W)\equiv (i\gamma\cdot\partial+W)\Delta_F(x|W)$, we will obey
gauge covariance by stipulating that, in any other gauge $M$, the
expressions above are to be multiplied by $\exp[ie^2M(x-y)]$. This then
leads to the gratifying result that the resulting vacuum polarization is
transverse and gauge-independent, since
\begin{eqnarray}
\Pi_{\mu\nu}(z) &\propto& \lim_{x,y\rightarrow 0}{\rm Tr}
                        (S\Gamma_\mu S)(x,y;z)\gamma_\nu\nonumber\\
 & = & \lim_{x,y\rightarrow 0}{\rm Tr}\int dW\,\rho_0(W)S_F(x-z|W)\gamma_\mu
         S_F(z-y|W)\gamma_\nu{\rm e}^{ie^2M(x-y)}\nonumber \\
 & = & {\rm Tr}\int dW\,\rho_0(W)S_F(x-z|W)\gamma_\mu S_F(z-y|W)\gamma_\nu,
\end{eqnarray}
if we approach the equal location limit $x=y$ along a certain direction.
This would imply that vacuum polarization is gauge independent and is given
by the Landau gauge result, $\Pi_{\mu\nu}(z) =\int dW\,\rho_0(W)
\Pi_{\mu\nu}(z|W)$, corresponding to a weighted mass integral. Actually this
result is still not correct: the Green function (14) is insufficient to
account
for all higher order quantum corrections, because we must supplement it by a
transverse (Landau gauge) contribution, which certainly does not spoil the
transversality property. Unless this is added, $\beta_{\rm QED}$, will
almost
certainly be wrong.

Thus, from all of this discussion, we see that the most pressing problem in
patching up the gauge technique is to incorporate transverse Green functions
which have correct analytic and gauge-covariance properties. Until this is
done, any {\em physical} results that are claimed to be a consequence of the
technique are not to be fully trusted.

\subsection*{Acknowledgements}
This work was initiated during the QFT98 workshop (Feb. 1998, University
of Adelaide) and I wish to thank the organizers, A Schreiber, A Thomas
and A Williams, for their hospitality there.

\section*{Appendix}
In scalar electrodynamics, if $\Delta(p^2)=\int dW^2\rho(W^2)/(p^2-W^2)$
stands for the meson propagator,
the longitudinal expressions for the vertex,
$$\Gamma^L_\mu(p',p)=(p+p')_\mu\frac{\Delta^{-1}(p'^2)-\Delta^{-1}(p^2)}
                 {p'^2-p^2}, \qquad {\rm and} $$
\begin{eqnarray*}
(\Delta\Gamma^L_\mu\Delta)(p',p)&=&(p+p')_\mu\frac{\Delta(p^2)
        -\Delta(p'^2)}{p'^2-p^2}\\
 &=& \int dW^2\rho(W^2)\left[\frac{1}{p'^2-W^2}(p+p')_\mu\frac{1}
             {p^2-W^2}\right],
\end{eqnarray*}
are entirely equivalent. Although one may add any amount of transverse
amplitude without affecting the WGT identity, it would be perverse to
introduce such an additional piece {\em unless there are good reasons to
do so} like reaching agreement with perturbation theory,
eliminating subdivergences or trying to patch up the gauge covariance
relation. This is precisely what has motivated King (in QED) and Haeri
(in QCD) to incorporate particular transverse terms in the ultraviolet
regime (King 1983; Haeri 1988). No such corrections are needed in the
infrared regime, since all Green functions are effectively governed by the
nonperturbative behaviour of the charged particle propagator (Delbourgo
and West 1977a, 1977b; Delbourgo 1979; Atkinson and Slim 1979).

In spinor electrodynamics the situation is much less clearcut because
of the plasticity of the gamma-matrix algebra.
Begin with the spinor propagator, written in the equivalent forms,
$$S(p)=\int dW \rho(W)/(\not{p}-W)\quad {\rm or}\quad S^{-1}(p) \equiv
       \not{p}A(p^2) + B(p^2).$$
For short, write $A\equiv A(p^2), B\equiv B(p^2), A'\equiv A(p'^2),
B'\equiv B(p'^2)$ and $F\equiv p^2A^2 - B^2$, etc. Then there are at
least three `obvious' ways of `solving the gauge identities', all of which
are singularity free. From the proper vertex identity, one may factor out
the momentum transfer and arrive at the first version (Ball and Chiu 1980),
$$\Gamma^L_\mu(p',p) = \frac{1}{2}\gamma_\mu(A'+A) +
  \frac{(p'+p)_\mu}{p'^2-p^2}\left[(B'-B)+\frac{1}{2}(\not{p}'+\not{p})
   (A'-A)\right]. $$
A second way is to carry out the factorization for the full Green function:
$$-(S\Gamma^L_\mu S)(p',p)=\frac{\gamma_\mu}{2}\left(\frac{A'}{F'}\!+\!
        \frac{A}{F}\right)+\frac{(p'+p)_\mu}{p'^2-p^2}\left[
\frac{\not{p}'\!+\!\not{p}}{2}\left(\frac{A'}{F'}\!-\!\frac{A}{F}\right)
        -\left(\frac{B'}{F'}\!-\!\frac{B}{F}\right)\right].
$$
A third way is to take advantage of the dispersive representation as a
weighted mass integral and thereby arrive at the form
$$ (S\Gamma^L_\mu S)(p',p) = \int dW\rho(W)\frac{1}{\not{p}'-W}
       \gamma_\mu \frac{1}{\not{p}-W}. $$
These three versions are not identical to one another, in contrast to
the scalar case. They differ from one another by particular transverse
components (which of course have no effect on the WGT identity) and
no version is more natural than any other at this level, unless other
considerations intervene; thus they all behave smoothly as $p^2
\rightarrow p^2$ and they agree with lowest order perturbation theory.
For instance, the difference between the first and second versions
of the proper vertices can be expressed as
$$\frac{A'}{(\not{p}'A'-B')}T_\mu(\not{p}A+B)+A'T_\mu +
  (\not{p}'A'+B')T_\mu\frac{A}{(\not{p}A-B)} + T_\mu A$$
where $2T_\mu=\gamma_\mu-(\not{p} '-\not{p})(p'+p)_\mu/(p'^2-p^2)$
is a transverse Lorentz-covariant. Similarly, the third version can be
rewritten in a more revealing form:
$$(S\Gamma^L_\mu S)(p',p)=\int dW\rho(W)\frac{1}{\not{p}'-W}2T_\mu
                \frac{1}{\not{p}-W}+\frac{(p'+p)_\mu}{p'^2-p^2}
                \left[S(p)-S(p')\right].$$
Haeri and Haeri (1992) have shown that this particular spectral form
of the longitudinal vertex can be converted into the equivalent but
more elegant form
$$(S\Gamma^L_\mu S)(p',p)=\frac{(\not{p}'\gamma_\mu+\gamma_\mu\not{p})
       S(p)-S(p')(\not{p}'\gamma_\mu+\gamma_\mu\not{p})}{p'^2-p^2}, $$
which corresponds to the proper vertex solution
$$\Gamma_\mu^L(p',p)=\frac{\not{p}'\gamma_\mu\not{p}(A'-A) +
(\not{p}'\gamma_\mu+\gamma_\mu\not{p})(B'-B)+\gamma_\mu(p'^2A'-p^2A)}
 {p'^2-p^2},$$
featuring the inverse propagator functions $A$ and $B$.

\subsection*{References}

{\small

\noindent Atkinson, D., and Slim, H. (1979). {\em Nuovo Cim.} {\bf 50A},
555.

\noindent Alekseev, A. I. (1998). Proceedings of Workshop on
``Nonperturbative
Methods in Quantum Field Theory'', p.104, ed. Schreiber, Williams and
Thomas,
Adelaide (World Scientific, 1998).

\noindent Ball, J. S., and Chiu, T-W. (1980). {\em Phys. Rev.} {\bf 22},
2542.

\noindent Cahill, R. T., and Gunner, S. M. (1998). Proceedings of Workshop
on
``Nonperturbative Methods in Quantum Field Theory'', p.261, ed. Schreiber,
Williams and Thomas, Adelaide (World Scientific, 1998).

\noindent Cornwall, J. M. (1986). {\em Phys. Rev.} {\bf D34}, 585.

\noindent Curtis, D. C., and Pennington, M. R. (1993). {\em Phys. Rev.}
{\bf D48}, 4933.

\noindent Delbourgo, R. (1979). {\em Nuovo Cim.} {\bf 49A}, 484.

\noindent Delbourgo, R., and West, P. C. (1977a). {\em J. Phys.} {\bf A10},
1049.

\noindent Delbourgo, R., and West, P. C. (1977b). {\em Phys. Lett.}
{\bf 72B}, 86.

\noindent Delbourgo, R., and Thompson, G. (1982). {\em J. Phys.}
{\bf G8}, L185.

\noindent Delbourgo, R., and Zhang, R. B. (1984). {\em J. Phys.} {\bf 17A},
3593.

\noindent Green, H. S. (1953). {\em Proc. Phys. Soc. (London)} {\bf A66},
873.

\noindent Haeri, B. J (1988). {\em Phys. Rev.} {\bf D38}, 3799.

\noindent Haeri, B. J. (1993). {\em Phys. Rev.} {\bf D48}, 5930.

\noindent Haeri, B. J., and Haeri, M. B. (1992). {\em Phys. Rev.}
{\bf D43}, 3732.

\noindent Ivanov, M. A., Kalinovskii, Yu. L., Maris, P., and Roberts, C. D.
(1998). 11th Int. Conf. of Problems in QFT, Dubna (nucl-th/9810010).

\noindent Johnson, K., and Zumino, B. (1959). {\em Phys. Rev. Lett.}
{\bf 3}, 351.

\noindent King, J. E. (1983). {\em Phys. Rev.} {\bf D27}, 1821.

\noindent Kizilersu, A., Reenders, M., and Pennington, M. R. (1995). {\em
Phys.
Rev.} {\bf 352}, 1242.

\noindent Kondo, K-I. (1997). {\em Int. J. Mod. Phys.} {\bf A12}, 5651.

\noindent Landau, L. D., and Khalatnikov, I. M. (1956). {\em Zh. Eksp.
Teor. Fiz.} {\bf 29}, 89 [{\em Sov. Phys. JETP} {\bf 2}, 2].

\noindent Larin, S. F., and Vermaseren, J. A. M. (1993). {\em Phys. Lett.}
{\bf B303}, 334.

\noindent Maris, P., and Roberts, C. D. (1997). {\em Phys. Rev.} {\bf56C},
3369.

\noindent Maris, P., and Roberts, C. D. (1998). Proceedings of Workshop
on ``Nonperturbative Methods in Quantum Field Theory'', p.132, ed.
Schreiber, Williams and Thomas, Adelaide (World Scientific, 1998).

\noindent Papavassiliou, J., and Cornwall, J. M. (1991). {\em Phys. Rev.}
{\bf D44}, 1285.

\noindent Parker, C. N. (1984). {\em J. Phys.} {\bf 17A}, 2873.

\noindent Roberts, C. D., and Williams, A. G. (1994). {\em Prog. Part. Nucl.
Phys.} {\bf 33}, 477.

\noindent Salam, A. (1963). {\em Phys. Rev.} {\bf 130}, 1287.

\noindent Salam, A., and Delbourgo, R. (1964). {\em Phys. Rev.} {\bf 135},
B1398.

\noindent Strathdee, J. (1964). {\em Phys. Rev.} {\bf 135}, B1428.

\noindent Takahashi, Y. (1957). {\em Nuovo Cim.} {\bf 6}, 371.

\noindent Thompson, G. (1983). {\em Phys. Lett.} {\bf 131B}, 385.

\noindent Thompson, G., and Zhang, R. B. (1987). {\em Phys. Rev.} {\bf D35},
631.

\noindent Ward, J. C. (1950). {\em Phys. Rev.} {\bf 78}, 182.

\noindent Zumino, B. (1960). {\em J. Math. Phys.} {\bf 1}, 1.
}

\end{document}